# Thermopower-enhanced efficiency of Si/SiGe ballistic rectifiers

D. Salloch[a)], U. Wieser, and U. Kunze,
*Werkstoffe und Nanoelektronik, Ruhr-Universität Bochum, D-44780 Bochum, Germany*

T. Hackbarth
*DaimlerChrysler Forschungszentrum Ulm, Wilhelm-Runge-Straße 11, D-89081 Ulm, Germany*

Injection-type ballistic rectifiers on Si/SiGe are studied with respect to the influence of gate voltage on the transfer resistance $R_T$ (output voltage divided by input current) for different positions of a local gate electrode. The rectifiers are trifurcated quantum wires with straight voltage stem and oblique current-injecting leads. Depending on the gate configuration, thermopower contributions arise from nearly-pinched stem regions which either cancel each other or impose upon the ballistic signal with same or opposite polarity. At best, this enhances $R_T$ to a maximum value of 470 Ω close to threshold voltage.

Ballistic full-wave rectification is a device functionality observed in asymmetric nanoscale three-[1,2] or four-terminal[3-7] cross junctions in the nonlinear response regime. Unlike classical rectifiers like *pn* junctions and Schottky contacts or nanodevice concepts exploiting field-effect induced depletion barriers,[8] the ballistic rectifier is free from any potential barrier and, hence, intrinsically exhibits a zero cut-in voltage. In three-terminal Y-[1] or T-shaped[2] devices the rectification effect mainly relies upon the change of the number of participating transport modes in the current-carrying leads. A different mechanism is effective in four-terminal cross junctions, where the ballistic electrons are guided into one of the current-free voltage leads independent of the input current direction. As a result this lead is stationary charged and, hence, a nonlocal voltage arises. Controlling the electron trajectories has been accomplished by a single asymmetrically shaped[3,4] or positioned[5] antidot scatterer in the current path of an orthogonal cross junction (reflection-type rectifier). Recently we have demonstrated the trajectory control by a non-centrosymmetric nanoscale cross junction, where the current-injecting leads are inclined[6] or parallel[7] with respect to the voltage leads (injection-type rectifier).

In the present work, we investigate the tunability of the injection-type rectifier by a local gate electrode with respect to an improved transfer resistance $R_T = |V_{out}/I_{in}|$, where $V_{out}$ is the output voltage and $I_{in}$ the input current. Figure 1 shows the device geometry with 220 nm wide stem and injection leads merging under an angle of 45° into the stem. In sample 1, instead of a global metal gate electrode[6] a local gate is positioned directly above the stem [Fig. 1(a)]. Two additional samples of identical rectifier geometry are equipped with local gates which only cover the crossing region and the adjacent part of the upper stem [sample 2, Fig. 1(b)] or the lower (collector) part of the stem [sample 3, Fig. 1(c)]. In samples 1 and 2 a negatively biased gate induces a density gradient between the injector and the stem electron system. Such a density step in a two-dimensional electron gas (2DEG) is known to cause a trajectory refraction analog to Snell's law as $\sin(\vartheta_1)/\sin(\vartheta_2) = (n_2/n_1)^{1/2}$, where $\vartheta_1$, $\vartheta_2$ denote the angles between trajectories and normal incidence and $n_1$, $n_2$ the densities in the regions 1 and 2, respectively.[9] For electrons crossing the boundary between the injector (region 1) and the stem (region 2) the effective injection angle can be reduced at an optimum gate voltage from the geometric value of 45° down to 0°. Further lowering the stem electron density leads to an electron reflection; this should result in a less effective electron injection. For comparison, a negative gate voltage in sample 3 creates the density step inside the stem, i.e. the collector part of the stem can be gradually depleted.

The samples are prepared from a modulation-doped Si/Si$_{0.7}$Ge$_{0.3}$ heterostructure with the 2DEG located 36 nm below the surface. The mobility and electron density are determined at 4.2 K from a Hall bar device and were found to be $\mu \sim 1.4 \times 10^5$ cm$^2$/Vs and $N_S \sim 5 \times 10^{11}$ cm$^{-2}$, respectively, leading to a mean free path of $l \sim 1.6\,\mu$m. The large-area reservoirs and the nanoscale cross-junction are patterned in subsequent processes of standard photolithography and electron-beam lithography forming a combined etch mask, followed by a single step of 50 nm deep low-damage reactive

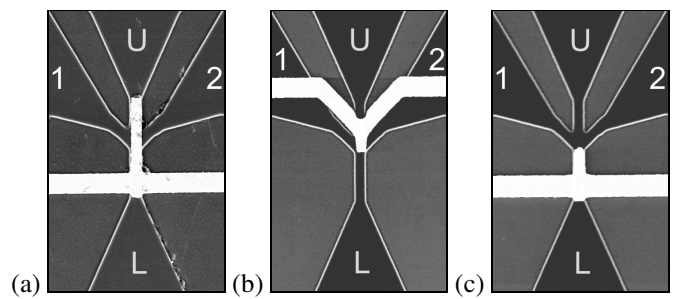

(a)  (b)  (c)

FIG. 1. Top-view scanning electron micrograph of the central part of asymmetric Si/SiGe cross junction. (1), (2) denote the contacts to the 220 nm wide and 600 nm long injectors, and (U) and (L) the upper and lower contact to the 220 nm wide and 2 μm long stem. The angle between the injectors and the stem is 45°. Metal gate electrodes (white) cover (a) the whole stem in sample 1, (b) the crossing and upper part of the stem in sample 2, and (c) the 1 μm long lower part of the stem in sample 3.

[a)]Electronic mail: daniel.salloch@rub.de

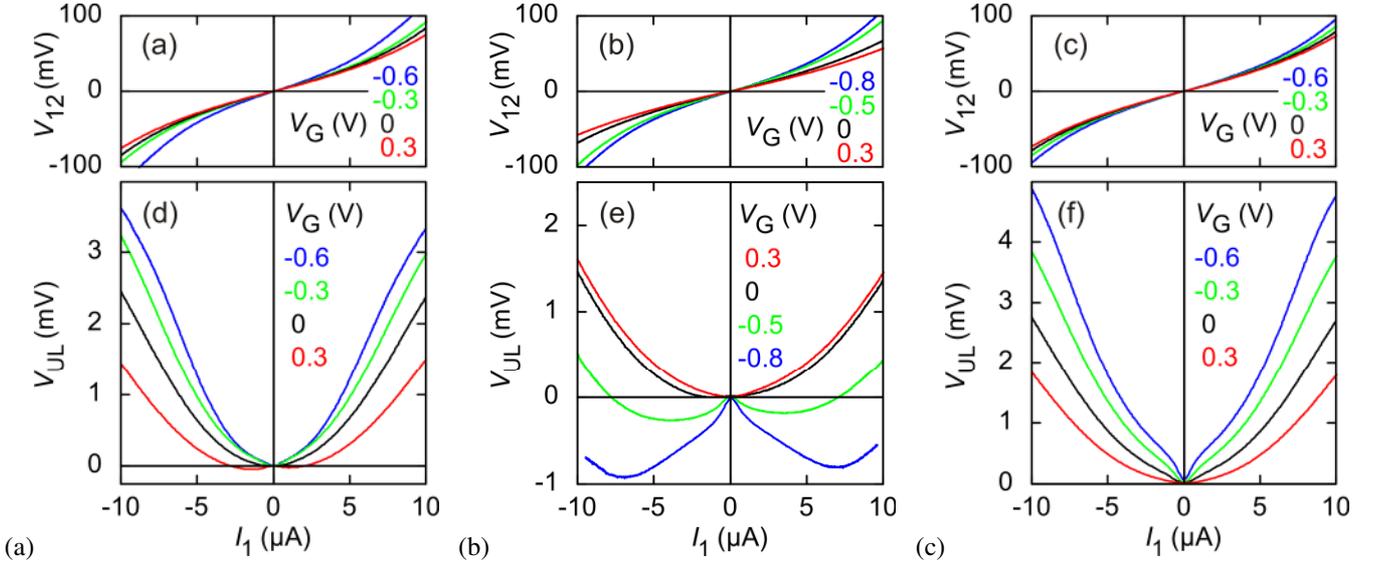

FIG. 2. (Color online) (a)–(c) Input characteristic $V_{12}$ versus $I_1$ and (d)–(f) transfer characteristic $V_{UL}$ versus $I_1$ of respectively (a), (d) sample 1, (b), (e) sample 2 and (c), (f) sample 3 at different gate voltages $V_G$. Note the different scales of the $V_{UL}$ axes. Gate leakage current at negative $V_G$ causes a $V_{UL}$ offset voltage. ($T$=4.2 K.)

ion etching.[10] Ohmic contact implantation of phosphorous ions was performed at an energy and dose of 30 keV and $2 \times 10^{15}$ cm$^{-2}$, respectively. Recrystallization of the amorphous surface layer and donor activation was achieved by annealing the sample at 560 °C for 30 min. Ti–Au contacts are prepared by optical lithography, metal evaporation, and lift off, followed by sintering at 400 °C for 1 min. Finally, the gate electrodes are formed by electron-beam lithography, Ni–Au evaporation and lift off. One chip containing two devices was assembled into a chip carrier, wire bonded and mounted into a He cryostat. Transport measurements were performed in a push-pull fashion with the driving voltage $+V_{12}/2$ applied at contact (1) and $-V_{12}/2$ at contact (2), where $V_{12} = V_1 - V_2$. The rectified signal $V_{out} = V_{UL} = V_U - V_L$ and the input current $I_{in} = I_1$ were recorded as a function of $V_{12}$ under fixed gate voltage $V_G$. The gate current was monitored in the same process.

The voltage-current ($V_{12}$-$I_1$) input characteristics of the samples are shown in Fig. 2(a)–(c). As can be seen the gate voltage has little effect on the channel current. This is the result of the gate electrode position. None of the local gates cover the narrow part of the injector channel; hence when the gated narrow part of the stem is pinched off the wide part is still conducting. The deviation from linearity at voltages $|V_{12}| > 30$ mV results mainly from carrier heating. The $V_{LU}$-versus-$I_1$ transfer characteristics with $V_G$ as parameter, Figs. 2(d)–2(f), are roughly parabolic for all samples at zero and positive gate voltage. We attribute the rectification effect to the inertial-ballistic injection of electrons into the collector part of the stem and the forming of a stationary dipole.[6]

At zero gate voltage the transfer characteristics of the samples 1, 2 and 3 are similar reaching at $I_1 = \pm 10$ μA $V_{LU} \sim 2.4$ mV, 1.4 mV, and 2.7 mV, respectively, which corresponds to a transfer resistance of $R_T = |V_{LU}/I_1| \sim 240$ Ω, 140 Ω, and 270 Ω. The magnitude of gate leakage currents is increased up to a maximum of 160 nA for gate voltages close to the stem threshold voltage. Note that only that fraction of the leakage current which flows from the gate electrode into the collector part of the stem gives rise to an offset voltage in $V_{UL}$ as shown in Fig. 2(f).

At negative gate bias the characteristics are completely different due to heating of the electrons in the electric field associated with the injection current. These hot electrons have enough energy to overcome the saddle-point potential at the constrictions formed in the upper and/or lower part of the stem. After energy relaxation of the electrons the constrictions are nearly blocking and thus create a voltage drop across the constrictions, which commonly is called hot-electron thermopower.[11] At positive and moderate negative gate voltage where the constrictions are conducting the thermovoltage is at least one order of magnitude smaller than the inertial-ballistic signal. Only closely above the threshold voltage the thermopower reaches appreciable values.[11] In sample 1 the gate electrode covers both constrictions. Owing to the position of the heating channel between the constrictions their thermo-voltages have opposite sign and should roughly cancel each other. In sample 2 only the upper stem constriction contributes a thermopower signal which is subtracted from the inertial-ballistic voltage. At low current the thermopower dominates and the output voltage changes its sign. About zero current the thermopower forms a narrow parabolic characteristic, while at high current the thermopower tends to saturate. This is also visible from the characteristic of sample 3 where the ballistic and thermoelectric contributions are imposed at equal polarity. The resulting output signal enhancement is of practical interest because not only the ballistic rectification but also the thermopower is an intrinsically fast process.

The effect of trajectory refraction may be a second source for improvement. In principle, the gate geometry of samples 1 and 2 should favor a contribution from refraction. However, due to the large thermopower in sample 2 it is not possible to extract such a contribution. Therefore we compare the gate-voltage dependent rectification efficiency of samples 1 and 3. Assuming $\vartheta_1 = 45°$ and $\vartheta_2 = 90°$, optimum injection into a pure 2DEG is attained for $n_2 = n_1/2$. If we regard the 220 nm wide stem as 2DEG, a refraction-induced optimum injection should arise at moderate negative gate voltage. In sample 1 where the thermopower has minor influence on the characteristic, optimum refraction is expected to create a maximum in $R_T$ as a function of $V_G$. This dependence is plotted in Fig. 3(a) at a constant current of $I_1 = 2\,\mu A$, $6\,\mu A$, and $10\,\mu A$. Prominent features are a magnitude, which is roughly proportional to the current, the maximum of $R_T$ close above threshold voltage, and a weak oscillation imposed upon the characteristic. For sample 3, Fig. 3(b), $R_T$ is less current-dependent, has the same maximum position, and again exhibits a weak oscillation. Both the maximum at threshold voltage and the oscillations can be attributed to the thermovoltage of a one-dimensional (1D) electron system.[11] Thus the essential difference between the samples is the magnitude of the thermopower contribution, which causes both a large voltage at low current close to the threshold voltage as well as an oscillatory behavior reflecting the 1D subband structure.[11] The lack of a major contribution from trajectory refraction in sample 1 is probably due to the 1D transport character.

The value of $R_T$ can be read as a useful figure of merit for ballistic rectification. In contrast to the voltage efficiency which has been used before[6] to benchmark the ballistic rectification, $R_T$ is not affected by the series resistance of the contacts and leads. In a comparative rating it should be considered that, due to the nonlinear $V_{LU}$-versus-$I_1$ transfer characteristic, $R_T$ depends on the driving current $I_1$. For possible device applications in low-level AC signal detection $I_1$ should be as low as possible. The best result of this work has been achieved in sample 3 as $R_T \approx 370\,\Omega$ at $I_1 = 2\,\mu A$ and $R_T \sim 470\,\Omega$ at $I_1 = 10\,\mu A$. The demonstration of ballistic rectification by an asymmetric scatterer was performed at $R_T \sim 5\,\Omega$ for $I_1 = 10\,\mu A$.[3] Injection-type rectifiers in AlGaAs/GaAs[6] and Si/SiGe[7] yielded values of $R_T \sim 120\,\Omega$ and $R_T \sim 110\,\Omega$ at $I_1 = 10\,\mu A$, respectively. In a reflection-type Si/SiGe rectifier with antidot scatterer ballistic rectification of reversed sign was observed at low currents,[7] which we quantify as $R_T \sim 180\,\Omega$ at $I_1 = 2\,\mu A$. In view of the large difference in the mean free path for electrons in the 2DEG between AlGaAs/GaAs ($l \sim 8\,\mu m$)[6] and Si/SiGe ($l \sim 1.6\,\mu m$) we deduce that the geometry of the device is most crucial for the device operation as ballistic rectifier.

In conclusion, depending on the gate position thermopower is an important contribution in addition to the inertial-ballistic mechanism in determining the magnitude and tunability of the transfer resistance of injection-type ballistic rectifiers. At least in cross junctions formed from quantum wires there is no clear signature of trajectory refraction by gate-controlled regions of different electron density. Therefore the controlled region can be reduced to the collector part of the voltage stem. As a result of an advantageous geometry at low temperatures the transfer resistance can be attained as high as $R_T \sim 470\,\Omega$.

This work was supported by the Deutsche Forschungsgemeinschaft and by the Ruhr-University Research School funded by Germany's Excellence Initiative [DFG GSC 98/1].

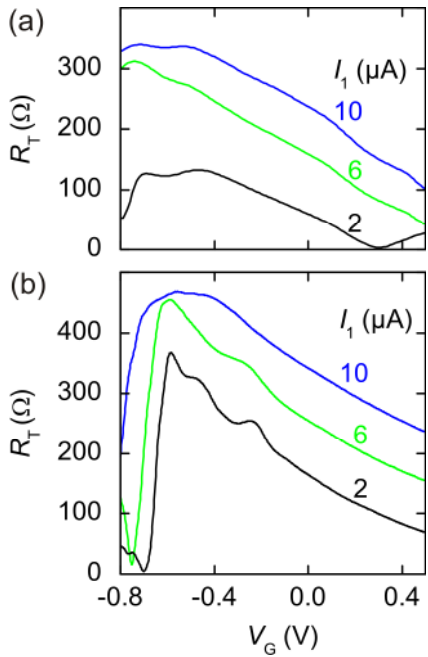

FIG. 3. (Color online) Transfer resistance $R_T = |V_{UL}/I_1|$ as a function of gate voltages $V_G$ of (a) sample 1, (b) sample 3 at different current $I_1$. ($T$=4.2 K.)